\documentclass[apjl]{emulateapj}
\usepackage{natbib}
\usepackage{apjfonts}
\usepackage{epsfig}

\newcommand{\mAB}{m_{\mathrm{AB}}}

\newcommand{\sigmaDM}{\sigma_{\mathrm{V}}}
\newcommand{\sigmaCV}{\sigma_{\mathrm{CV}}}
\newcommand{\ave}[1]{\langle #1 \rangle}

\shorttitle{COSMIC VARIANCE IN THE FRONTIER FIELDS}
\shortauthors{ROBERTSON ET AL.}

\begin{document}

\title{Accounting for Cosmic Variance in Studies of Gravitationally-Lensed High-Redshift Galaxies in the Hubble Frontier Field Clusters}

\author{Brant E. Robertson\altaffilmark{1}, Richard S. Ellis\altaffilmark{2}, James S. Dunlop\altaffilmark{3}, Ross J. McLure\altaffilmark{3}, Dan P. Stark\altaffilmark{1}, Derek McLeod\altaffilmark{3}}
\altaffiltext{1}{Steward Observatory, University of Arizona, 933 North Cherry Avenue, Tucson, AZ 85721, USA}
\altaffiltext{2}{Department of Astronomy, California Institute of Technology, MS 249-17, Pasadena, CA 91125, USA}
\altaffiltext{3}{Institute for Astronomy, University of Edinburgh, Royal Observatory, Edinburgh EH9 3HJ, UK}

\begin{abstract}
Strong gravitational lensing provides a powerful means for studying faint galaxies in the
distant universe. By magnifying the apparent brightness of background sources, massive clusters
enable the detection of galaxies fainter than the usual sensitivity limit for blank fields.
However, this gain in effective sensitivity comes at the cost of a reduced
survey volume and, in this {\it Letter}, we demonstrate there is an associated increase in
the cosmic variance uncertainty. As an example, we show that the cosmic variance uncertainty 
of the high redshift population viewed through the Hubble Space Telescope Frontier Field
cluster Abell 2744 increases from $\sim35\%$ at redshift $z\sim7$ to $\gtrsim65\%$
at $z\sim10$. Previous studies of high redshift galaxies identified in the Frontier Fields
have underestimated the cosmic variance uncertainty that will affect the ultimate
constraints on both the faint end slope of the high-redshift luminosity function and 
the cosmic star formation rate density, key goals of the Frontier Field program.
\end{abstract}


\section{Introduction}
\label{section:intro}

The remarkable capabilities of Wide Field Camera 3 (WFC3) on {\it Hubble Space
Telescope} (HST) have transformed infrared extragalactic surveys
of the distant universe. The Cosmic Assembly Near-infrared Deep 
Extragalactic Legacy Survey
\citep[CANDELS:][]{grogin2011a,koekemoer2011a}, the 
Cluster Lensing And Supernova survey with Hubble \citep[CLASH:][]{postman2012a}, and 
the Ultra Deep Field surveys 
\citep[UDF:][]{beckwith2006a,ellis2013a,koekemoer2013a,illingworth2013a}
have provided critical new information about the
rest-frame ultraviolet properties of early galaxies, their redshift-dependent abundance,
and the development of morphological structures over time
\citep[e.g.,][]{mclure2010a,oesch2010a,bouwens2011a,finkelstein2012a,schenker2013a,mclure2013a,dunlop2013a,ono2013a,curtis-lake2014a}.

The deepest HST observations to date in the UDF have reached multi-band sensitivities of 
$\mAB\approx29.5-30$ \citep[e.g.,][]{ellis2013a} after a total exposure
of hundreds of hours in a ``blank'' (i.e., devoid of strong lensing) field. 
To supplement the high-redshift galaxy populations discovered
in the UDF and its parallel fields, the currently on-going Frontier Fields program
(Program ID 13495; PI Lotz, Co-PI Mountain) utilizes carefully selected
strong gravitational lens clusters to probe intrinsically fainter limits
through high magnifications. With the ability to detect galaxies
with intrinsic magnitudes as faint as $\mAB\sim32$, the Frontier Fields program has
the potential to constrain the galaxy luminosity function faint-end
slope at redshifts $z>6$ and probe the UV luminosity density out to $z\sim12$.
Such constraints can provide vital clues to the process of cosmic reionization \citep{robertson2010a},
as previous analyses have suggested that the ionizing photon budget at
$z\sim7$ is dominated by faint galaxies below the current UDF limits 
\citep[e.g.,][]{robertson2013a}. Indeed, the first Frontier Fields observations of
the cluster Abell 2744 (A2744) have already been used to identify galaxy candidates in the 
reionization epoch \citep{atek2014a,atek2014b,zheng2014a,zitrin2014a,ishigaki2014a} 
and to constrain the luminosity density at redshift $z\sim10$ \citep{oesch2014a}. These results
complement discoveries of strongly-lensed high-redshift galaxies in the CLASH survey  
\citep{zheng2012a,coe2013a,bradley2014a}.

Utilizing lensed observations to infer constraints on the early galaxy
populations requires careful considerations of the volumes probed and the
associated uncertainties. This {\it Letter} presents the first estimates of the
{\it cosmic variance} of high-redshift galaxy samples in the Frontier Fields (FF) survey. Using
the publicly-available magnification maps for the first FF cluster, Abell 2744, we estimate 
the effective survey volume as a function of magnification and calculate 
the associated  cosmic variance uncertainty. Since the magnification varies significantly 
across a given cluster lens,  we use the connection between
magnification, effective survey volume, and cosmic variance uncertainty to produce a ``cosmic
variance map''. Importantly, in regions of extreme magnification, where the gain of
lensing is most valuable, the cosmic variance uncertainty is increased relative
to that for comparable blank-field surveys. This uncertainty
has important implications for the 
benefits of the FF program in its stated goals, as we attempt to quantify.

\begin{figure*}
\figurenum{1}
\begin{center}
\includegraphics[width=7.15in]{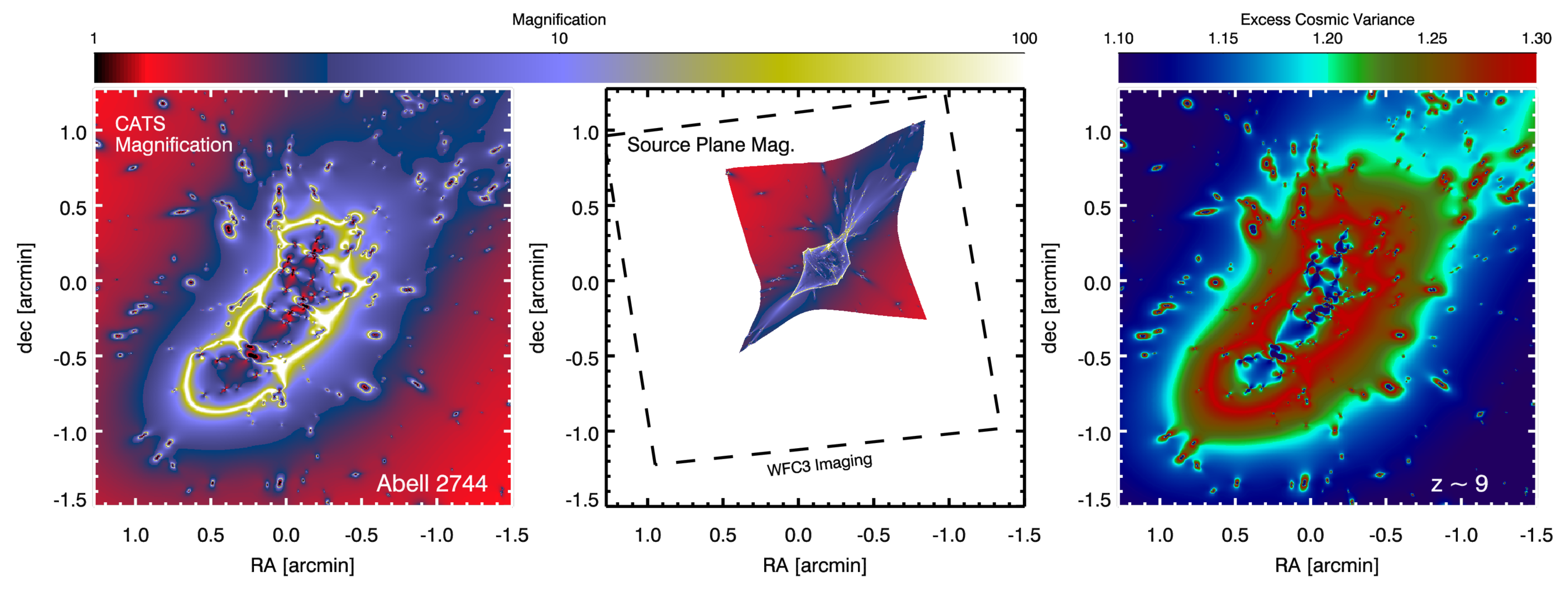}
\caption{\label{fig:maps}
Estimating the cosmic variance uncertainty for the Frontier Fields. The CATS Abell 2744 (A2744)
magnification map \citep[left panel; $z\sim9$;][]{richard2014a} shows the image plane 
amplification of flux from background sources caused by deflection from A2744. 
The corresponding deflection maps provided by \citet{richard2014a} can be used to 
recover the source plane magnification and effective survey area (middle panel, reconstructed 
for the observed A2744 
WFC3 field-of-view shown as a dotted line). The
cosmic variance uncertainty can then be estimated. This comparison provides the ``excess'' 
cosmic variance map of this lensed field over a blank field with the a same image area, 
assuming a constant bias population  (right panel, evaluated for a  $z\sim9$ sample). The
cosmic variance in this Frontier Field is $\sim10-30\%$ higher than for an equivalent blank
field high-redshift survey.
}
\end{center}
\end{figure*}

Throughout this {\it Letter} we adopt the flat $\Lambda$CDM cosmology 
($\Omega_m=0.3$, $\Omega_\Lambda=0.7$, $h=0.7$) used to produce the 
\citet{richard2014a} lensing maps of A2744. 
We further adopt the normalization of the linear power spectrum 
$\sigma_8 = 0.829$, spectral index $n=0.96$, and baryon density
$\Omega_b = 0.0487$ measured by \citet{planck2013a}.

\section{Luminosity-Dependent Cosmic Variance}
\label{section:cv}

The cosmic variance (CV) uncertainty of an observed galaxy population
reflects fluctuations in the matter density about the mean cosmic density, as sampled by 
the survey volume.  In linear theory, the galaxy number density $n$ in a volume will 
differ from the mean number density $\bar{n}$ as $n = \bar{n}(1+b\delta)$, 
where $\delta$ is the matter overdensity in the survey volume, and
$b$ is the clustering bias of the galaxy population. 

The bias $b$ and survey volume probed will in general depend on the
galaxy luminosity, which is important in the context of a strong lens survey 
where the effective volume varies strongly with intrinsic source flux.
In an unlensed blank field, the sample covariance matrix
$S_{ij} = \langle(n_i - \bar{n}_i)(n_j-\bar{n}_j)\rangle$ of the number of 
galaxies $n_i$ and $n_j$ in luminosity or magnitude bins $i$ and $j$ depends 
on the bias of the galaxy populations $b_i$ and $b_j$, and the average numbers of 
galaxies $\bar{n}_i$ and $\bar{n}_j$ expected in the survey \citep[for details see, e.g.,][]{robertson2010b,robertson2010c}.

The diagonal terms $S_{ii}$ of this matrix provide the
cosmic variance $\sigmaCV^2$ of the total galaxy number counts
typically expressed as a fractional uncertainty
\begin{equation}
\sigma_{CV} = \langle \sqrt{S_{ii}}/\bar{n}_i \rangle_{i} = \ave{b} \sigma_{DM} D(z),
\end{equation}
where $\langle\ldots\rangle_{i}$ denotes a suitable averaging of the luminosity-dependent
bias of the observed sample, and 
results in the product of an average bias $\ave{b}$, the growth factor $D(z)$,
and the rms matter density fluctuations $\sigma_{DM}$ in the survey volume at $z=0$ assuming
the effective survey geometry is luminosity-independent \citep[see Sections 2.3 and 2.4 of][]{robertson2010c}.
In the absence of direct clustering constraints, we estimate the bias $b$ by using abundance matching 
\citep{kravtsov2004a,conroy2006a} to assign dark matter masses to galaxies based on the
\citet{tinker2008a} halo mass function and then applying the bias model of 
\citet{tinker2010a}.

\section{Estimating Cosmic Variance in a Strongly-Lensed Survey}
\label{section:cv_lensed}

For a field with strongly varying magnification, the preceding calculation does not 
account for spatial variations in the range of intrinsic luminosities probed or the 
survey geometry as a function of magnification.  To model the covariance matrix in the 
strong lensing case, we consider a covariance matrix with a spatial dependence 
on the local magnification $\mu$ of the form
\begin{equation}
\label{eqn:lensed_covariance}
S_{ij}(\mu) = b_i b_j \bar{n}_i \bar{n}_j D^2(z) \int \frac{d^3 k}{(2\pi)^3} P(k) \hat{W}_i(\mathbf{k},\mu) \hat{W}_i^{\star}(\mathbf{k},\mu),
\end{equation}
\noindent
where $\hat{W}_i(\mathbf{k},\mu)$ describes the Fourier transform of the
subvolume of the survey with magnification $\mu$ as reconstructed in the source plane, and $P(k)$ is the
matter power spectrum \citep[e.g.,][]{eisenstein1998a}.

To estimate the sample variance $S_{ii}$ of a galaxy population with a range of
magnifications, some averaging is needed. For any intrinsic 
luminosity bin $i$, there exists a minimum magnification $\mu_{i}$ below which
the source flux will not be sufficiently amplified to be detected by the survey.
When the luminosity bin $i$ corresponds to a flux brighter than the 
nominal blank-field sensitivity of the survey, then sources of that intrinsic
brightness amplified by any magnification should be detected (i.e., $\mu_{i}=1$).
For intrinsically fainter objects, we have $\mu_{i}>1$. To estimate the CV
of objects in a luminosity bin $i$, we reconstruct the
source plane from a lens model and compute the effective source plane area of the survey 
$A(\mu>\mu_{i})$ with magnifications $\mu$ greater than $\mu_{i}$.
The integral over the power spectrum required to estimate the rms density fluctuations
$\sigmaDM$ in such an area can be evaluated using the window $\hat{W}(\mathbf{k})$
as in the blank-field case, but with an effective area $A(\mu>\mu_{i})$.
Regions within a survey with a given magnification $\mu$ can display a complicated topology, such that
evaluating $\hat{W}(\mathbf{k},\mu>\mu_{i})$ would prove difficult.
Instead, we model the source
plane area as a square. This choice has little impact since the line-of-sight
extent of the survey volume is much larger than its transverse size.

The remainder of the CV 
calculation then proceeds as described in
Section \ref{section:cv}, with the bias and rms density fluctuations probed by the 
luminosity-dependent effective survey volume averaged over luminosity and 
magnification to compute a characteristic CV 
$\ave{\sigmaCV} \approx  \ave{b} \ave{\sigmaDM}D(z)$.

\begin{figure}
\figurenum{2}
\includegraphics[width=3.3in]{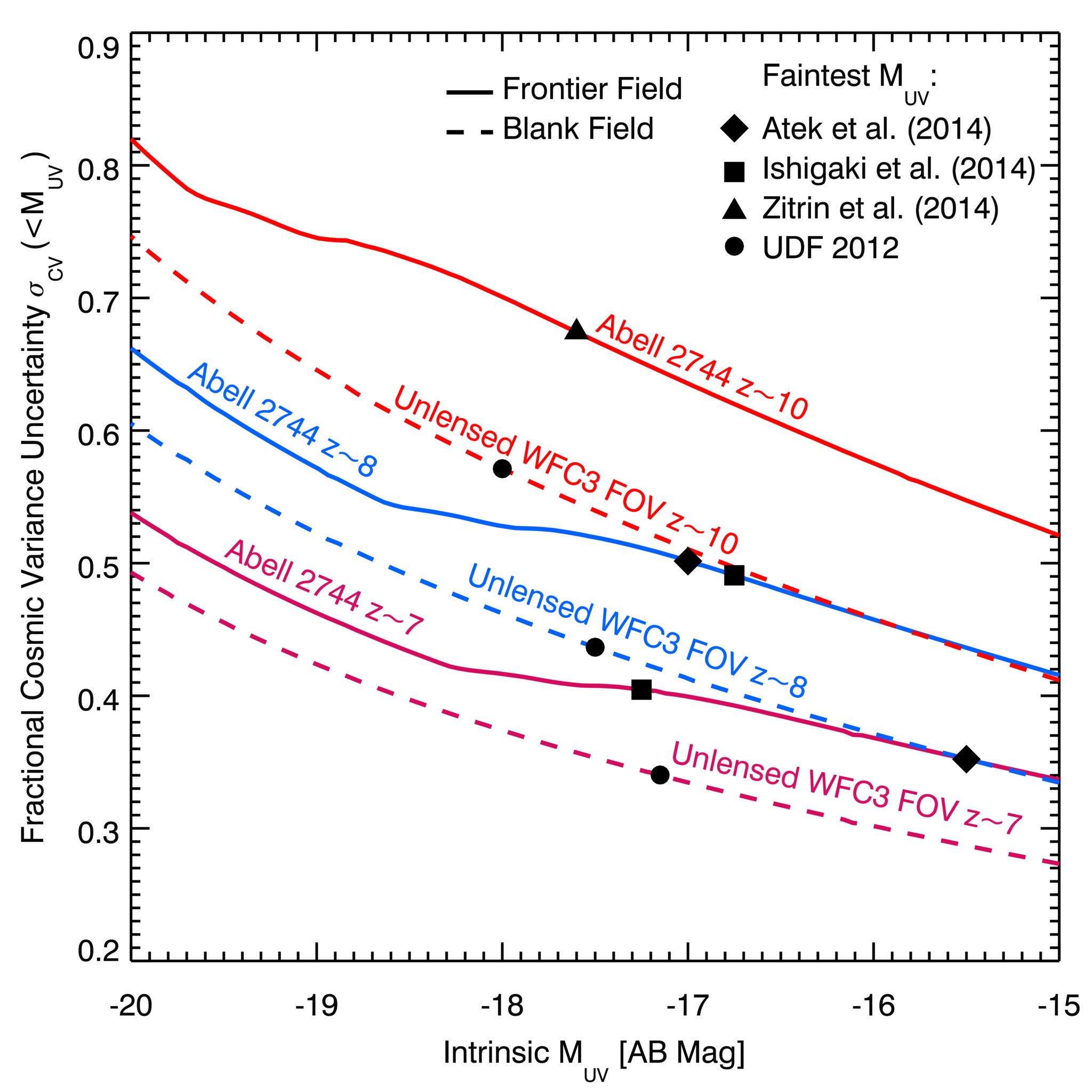}
\caption{\label{fig:cv} Fractional cosmic variance uncertainty in galaxy counts.
Cosmic variance in blank field surveys (dashed lines) can be estimated by computing 
the rms density fluctuations in the survey volume using linear theory and the luminosity-dependent 
clustering bias of galaxies from abundance matching (see Section
 \ref{section:cv}). Cosmic variance estimates for single WFC3 pointings are plotted at
 $z\sim7$ (magenta), $z\sim8$ (blue), and $z\sim10$ (red), along with the corresponding values
 for the UDF12 survey \citep[][points]{ellis2013a,schenker2013a,mclure2013a}. For strong gravitational
 lens surveys, the source plane area as a function of magnification can be used to determine a
 similar linear theory estimate of the cosmic variance in a lensed sample.  The corresponding
 cosmic variance uncertainty for A2744 
 is computed (solid lines) and indicated for the
 $z\sim7-8$ \citet[][diamonds]{atek2014b} and \citet[][squares]{ishigaki2014a} samples
  and $z\sim10$ \citet[][triangle]{zitrin2014a} object.
}
\end{figure}

\section{Cosmic Variance Uncertainties for the Frontier Fields}
\label{section:cv_ff}

Applying the methods presented in Sections \ref{section:cv} and \ref{section:cv_lensed}
to the Frontier Fields (FF) requires using magnification and deflection maps of individual cluster
lenses to reconstruct the effective area of the {\it HST} survey in the source plane.
Figure \ref{fig:maps} illustrates our methodology applied to A2744.
We use the Clusters As TelescopeS (CATS) lens models presented in \citet{richard2014a} 
that provide a map of the spatially-dependent
magnification (left panel of Figure \ref{fig:maps}, shown for the $z\sim9$ model).
The public \citet{richard2014a} models also include a matrix of deflections that allows for a 
reconstruction of a source plane magnification map. We use the {\it HST} WFC3  
weight map from the public FF data (Program ID 13495; PI Lotz, Co-PI Mountain)
to determine the area of A2744 covered by WFC3 imaging, and then reconstruct the
source plane magnification map of this region (our method is similar to that presented by
\citealt{coe2014a} and produces similar results to their Figure 5). The
reconstructed source plane magnification map is shown in the middle panel of Figure \ref{fig:maps},
and enables us to compute the area $A(\mu>\mu_i)$ that defines the intrinsic luminosity-dependent
window function used in Equation \ref{eqn:lensed_covariance} to calculate the sample variance.
The connection between magnification, source plane effective area, and CV can then
be used to produce a ``cosmic variance map'' of A2744. 
The right panel of Figure \ref{fig:maps}
shows the estimated excess CV 
in the A2744 field relative to a
blank field of the same imaging area, as a function of the local magnification. The CV
in A2744 is estimated to be 
$10-30\%$ higher than in an equivalent blank field survey, assuming a constant bias
population. 
Applying the same methodology to the other FF lens models suggests 
similarly increased uncertainties.

The luminosity-dependent CV 
uncertainty of the A2744 lens galaxy
population can be estimated as a function of intrinsic source flux.  Figure \ref{fig:cv}
shows the fractional CV uncertainty of the high-redshift
galaxy population statistics for unlensed surveys the size of a single
WFC3 field-of-view (dashed lines) and for a lensed population behind A2744 (solid
lines), calculated assuming the redshift-dependent luminosity function parameters
presented in \citet{bouwens2014a}.  
The CV 
uncertainty is computed for $z\sim7$ (magenta), $z\sim8$
(blue), and $z\sim10$ (red) populations.  We have additionally indicated the
CV estimates
for the UDF 2012 survey \citep{ellis2013a, schenker2013a, mclure2013a}, the \citet{atek2014b}
and \citet{ishigaki2014a} 
A2744 samples, and the \citet{zitrin2014a} $z\sim10$ object identified in the A2744
data. The A2744 samples have CV uncertainties
comparable to blank field surveys with depths $\sim2$ magnitudes
brighter. Since the CV of the lensed fields
depends mostly on the source plane 
effective area as a function of 
magnification, Figure \ref{fig:cv} should
provide a useful CV estimate for any FF
high-redshift sample.

\begin{figure}
\figurenum{3}
\includegraphics[width=3.4in]{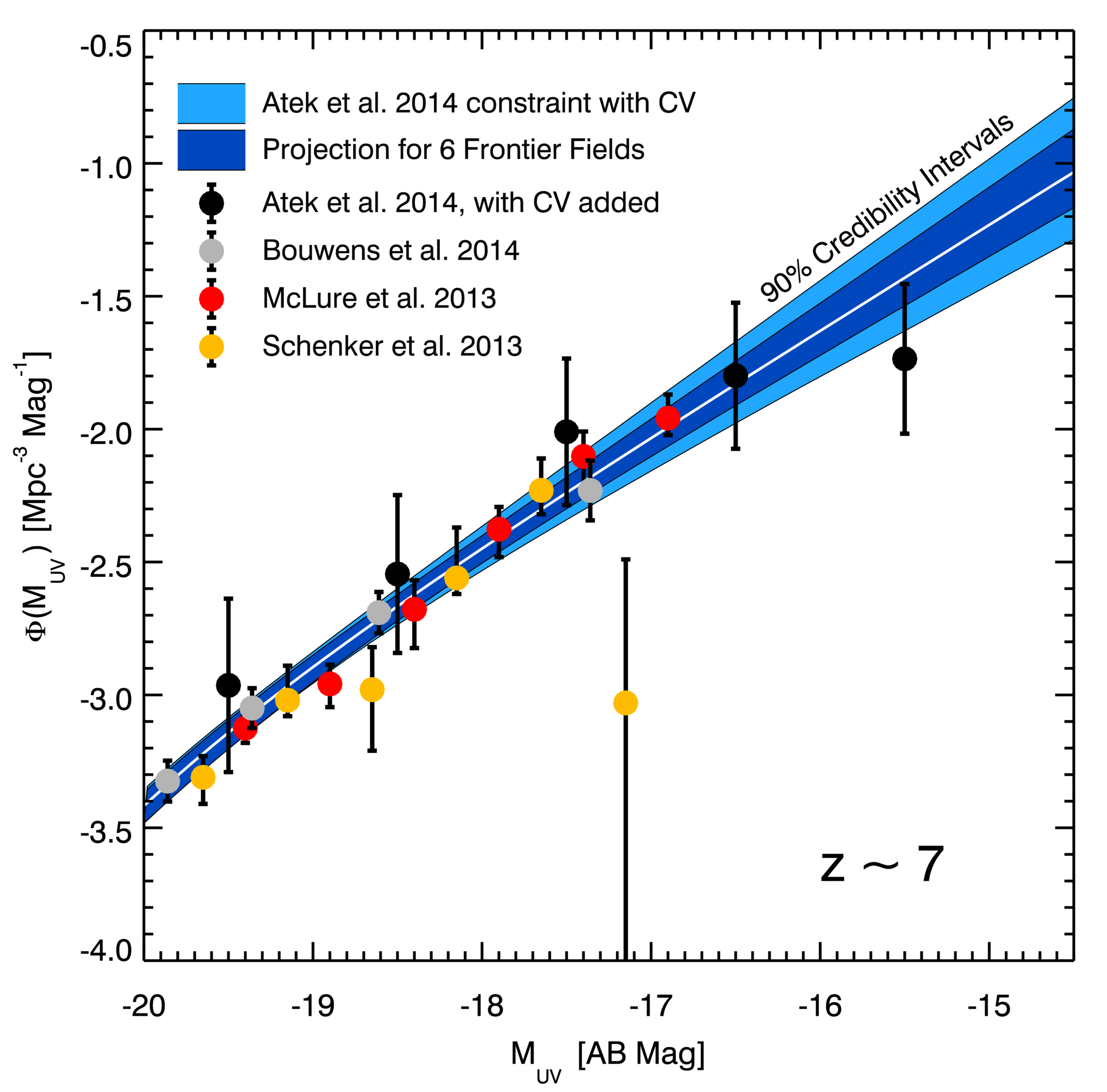}
\caption{\label{fig:lf} Revised $z\sim7$ luminosity function (LF) constraints from the
Abell 2744 (A2744) sample accounting for cosmic variance, and projections for constraints
from the full Frontier Fields program. Shown are the multi-field $z\sim7$ LF measurements 
from \citet[][gray points]{bouwens2014a}, and the A2744 measurements
from \citet[][black points]{atek2014b} with amplified error bars reflecting the newly
estimated cosmic variance uncertainty. The light blue region shows the 90\% credibility
intervals for the LF when constrained by the \citet{bouwens2014a} and modified
\citet{atek2014b} data. The \citet[][red points]{mclure2013a} and
\citet[][orange points]{schenker2013a} data are shown for 
comparison.
Assuming our best-fit LF parameters (white line) 
are accurate and A2744 is a representative lens, data from five additional clusters are 
simulated and used to project the constraints from the complete Frontier Fields program (dark blue area). 
When completed, we estimate that the full Frontier Fields program will deliver an uncertainty in the $z\sim7$
faint-end slope of $|\sigma_{\alpha}| \lesssim 0.05$.
}
\end{figure}

\section{Discussion}
\label{section:discussion}

{\it HST} Frontier Fields (FF) observations began in Cycle 21, and the program data has already 
identified distant galaxies behind A2744 
\citep{atek2014a, atek2014b, zheng2014a, zitrin2014a, oesch2014a}.
Several FF analyses have referred to the blank-field calculations of 
\citet{trenti2008a} to determine the CV of A2744 samples 
\citep[e.g.,][]{atek2014a,coe2014a,yue2014a}, but this model (and that discussed by \citealt{robertson2010c}) 
underestimates the CV 
uncertainty of gravitationally lensed populations.
\citet{zheng2014a} comment on the possibility of an increased 
CV for their sample owing
to lensing but provide no estimates. 
The new calculations presented in this
{\it Letter} account for the increased CV in the FF relative
to blank fields owing to the reduced effective volume 
of lensed surveys.\footnote{During the publication process, \citet{atek2014b} was revised to reflect our CV estimates.}

Understanding the CV of the FF
samples is critical for interpreting highly-magnified faint objects in the
broader context of the cosmic reionization process.
The robust identification of a handful of
extremely faint $z\sim7-8$ objects in the FF
could substantially improve the determination of the faint-end slope of the
high-$z$ luminosity function, as indicated by the sample of \citet{atek2014b} that
reaches down to $M_{UV}\sim-15$. The ionizing photon luminosity density provided by
high-$z$ galaxies identified above the limiting magnitude of the UDF ($M_{UV}\sim-17$ at $z\sim7$)
does not appear sufficient to
reionize the universe fully by $z\sim6$ under standard assumptions for the escape fraction and
ionizing photon production per unit UV luminosity \citep{robertson2013a}.
We infer that yet fainter
galaxies must provide a significant contribution to the UV luminosity density, and therefore 
our understanding of the
role of star-forming galaxies in reionization depends critically on 
uncertainties in the faint-end slope of the UV LF determination 
\citep{bolton2007a,robertson2010a,robertson2013a,kuhlen2012a}.
Among
the most precise determinations of the LF faint-end slope $\alpha$
at $z\sim7,8$ that fully accounts for the CV 
uncertainty of these 
faint, distant galaxy samples was provided by \citet{schenker2013a} using the
UDF and CANDELS Deep data, who found $\alpha(z\sim7) = -1.87^{+0.18}_{-0.17}$
and $\alpha(z\sim8) = -1.94^{+0.21}_{-0.24}$ \citep[see also][]{mclure2013a}. 
Similar faint-end slopes and uncertainties have been measured
independently 
\citep{oesch2012a,bouwens2014a} including using the 
A2744 sample
\citep{atek2014b}. As the lensed samples probe further down the luminosity function
with highly magnified objects, abundance matching suggests that
the clustering bias of the galaxy population is expected to decrease faster than the
reduced source plane effective volume causes the rms density fluctuations to increase.
Reaching substantially fainter galaxies therefore 
improves the CV statistics.

With an estimated CV uncertainty for the
A2744 sample, we can revisit the analysis 
presented by \citet{atek2014b}
accounting for CV 
and estimate the additional constraints that might be provided
by the complete FF program assuming A2744 is representative.
Figure \ref{fig:lf} shows the multi-field luminosity function
data from \citet{schenker2013a}, \citet{mclure2013a},
and \citet{bouwens2014a}, and 
the A2744 data from \citet{atek2014b}. We have increased the uncertainties
of the A2744 luminosity function data by adding the luminosity-dependent
CV uncertainty shown in Figure \ref{fig:cv} in 
quadrature with the errors reported by \citet{atek2014b}.
Performing Bayesian parameter estimation based on the {\it Multinest} sampling 
algorithm \citep{feroz2009a} and the \citet{bouwens2014a}
and \citet{atek2014b} data, 
we constrain the 90\% credibility interval for the $z\sim7$
luminosity function as shown in Figure \ref{fig:lf} (light blue area).  Assuming
our best-fit 
luminosity function parameters ($\phi_{\star} = 3.28\times 10^{-4}~\mathrm{Mpc}^{-3}~\mathrm{Mag}^{-1}$, 
$M_{\star}=-20.79$, $\alpha=-1.99$) are accurate and A2744
is a representative lens, we then perform
Monte Carlo realizations of the galaxy population in five additional FF
including the expected CV.
Repeating our parameter estimation on these bootstrapped models of the complete
six-cluster FF program (including the \citealt{bouwens2014a} data
as before) we find that the 90\% credibility interval on the luminosity function
shrinks considerably (dark area in Figure \ref{fig:lf}). Importantly, this result
suggests the complete FF program can provide critical 
information on the cosmic production rate of Lyman continuum photons by faint galaxies 
required to reionize the intergalactic medium by $z\sim6$. We forecast that the complete
FF program may reduce the uncertainty on the $z\sim7$ faint-end slope
to $\sigma_{\alpha} \lesssim0.05$ and the fractional uncertainty in 
UV luminosity density extrapolated to $M_{UV}=-13$ by a factor of $2\times$
to $\sim30\%$. The FF program may therefore help 
resolve whether star-forming galaxies were primarily responsible for completing the
cosmic reionization process. The FF may also help
constrain the evolution of the global star formation history at $z\sim7-10$, but 
such an analysis will require a careful treatment of the CV
of lensed
populations. 

We conclude by highlighting some features and limitations of our CV calculations for the FF
program. The computation of the source plane area requires the use of a lens
model and, while we use the CATS model of A2744 presented by \citet{richard2014a},
picking a different public lens model \citep[e.g.,][]{johnson2014a} can change the source plane
effective volume by $>10\%$ \citep[see Figure 5 of][]{coe2014a}.  The range of source plane effective
areas among the FF clusters is about a factor of 3, with A2744 being among the
largest.  The typical CV uncertainty of the high-redshift samples in the
other FF will be comparable to or slightly greater than that of A2744, provided
the intrinsic luminosity distributions of the sources are comparable.

The typical CV
uncertainty is of order unity, suggesting that our quasilinear model may underestimate
the true sample variance. The highly-lensed volumes are extremely small ($V\lesssim100$Mpc$^3$ for
magnifications $\mu\geq10$; see, e.g., Figure 5 of \citealt{coe2014a}), so nonlinear halo bias
may complicate the clustering statistics 
\citep[e.g.,][]{fernandez2012a,kitaura2014a}.
Precise applications of the FF samples for constraining
the luminosity function or high-redshift star formation rate 
density may therefore require more detailed modeling.

\section{Summary}
\label{section:summary}
The large
clustering bias of early galaxy populations and small
volumes probed by distant surveys make cosmic 
variance an important source of uncertainty for high-redshift
observations. These concerns are intensified for strongly-lensed
surveys like the Frontier Fields, as the amplification of source
fluxes through gravitational magnification comes at the cost of
a decreased effective survey volume. We present the first estimates of
the cosmic variance uncertainty associated with distant galaxy populations identified in the
Frontier Fields, using Abell 2744 as a representative example. By our 
estimates, the cosmic variance uncertainty increases from $\sim35\%$ for
the redshift $z\sim7$ sample of \citet{atek2014a,atek2014b} to $\gtrsim65\%$
for inferences drawn from the $z\sim10$ object examined by \citet{zitrin2014a}
and \citet{oesch2014a}. While these cosmic variance uncertainties are amplified
relative to blank-field surveys like the Ultra Deep Field \citep{beckwith2006a,ellis2013a},
they provide an independent sample to improve
luminosity function and star formation rate density estimates at high-redshift, provided
that their statistical properties are handled appropriately \citep{mcleod2014a}.

\acknowledgments
We thank Hakim Atek for providing tabulated data.
BER is supported in part by the National Science
Foundation under Grant No. 1228509, 
Grant No. NSF PHY11-25915 that funds the Kavli
Institute for Theoretical Physics at the University of California, Santa Barbara,
and by Space Telescope Science Institute under award HST-GO-12498.01-A. 
JSD acknowledges the support of the European Research
Council via the award of an Advanced Grant, and the
contribution of the EC FP7 SPACE project ASTRODEEP (Ref.No: 312725). RJM acknowledges ERC funding via the award of a consolidator grant (PI McLure).


\end{document}